\documentclass[preprintnumbers, prd, twocolumn, showpacs, floatfix, preprintnumbers, letterpaper, nofootinbib, amsmath, amssymb, superscriptaddress]{revtex4}
\usepackage{amssymb,amsmath,bm,dcolumn,epsf,graphicx,latexsym,slashed,simplewick}

\def\be{\begin{equation}}
\def\ee{\end{equation}}
\def\bea{\begin{eqnarray}}
\def\eea{\end{eqnarray}}

\usepackage{color}

\begin{document}

\title{New DBI Inflation model with Kinetic Coupling to Einstein Gravity}

\author{Taotao Qiu}
\affiliation{Key Laboratory of Quark and Lepton Physics (MOE) and College of Physical Science $\&$ Technology, Central China Normal University, Wuhan 430079, P.R.China}
\affiliation{State Key Laboratory of Theoretical Physics, Institute of Theoretical Physics, Chinese Academy of Sciences, Beijing 100190, China}

\pacs{98.80.Cq}

\begin{abstract}
In this letter we study a new class of inflation models which generalize the Dirac-Born-Infeld (DBI) action with the addition of a nonminimal kinetic coupling (NKC) term. We dubbed this model as the {\it new DBI inflation model}.
%The DBI-type action can be derived from more fundamental theories such as string theory. 
The NKC term does not bring new dynamical degree of freedom, so the equations of motion remain of second order. However, with such a coupling, the action is no longer linear with respect to the Einstein curvature term ($R$ or $G^{\mu\nu}$), which leads to a correction term of $k^4$ in the perturbations. The new DBI inflation model can be viewed as theories beyond Horndeski. Without violating nearly scale-invariance, such a correction may lead to new effects on the inflationary spectra that could be tested by future observations.
\end{abstract}
\maketitle

\section{Introduction} 
Recently there has been much effort in the study of generalized scalar field theories. Examples  of such non-canonical theories include the Galileons \cite{Nicolis:2008in, Deffayet:2009mn},  Horndeski theories \cite{Horndeski:1974wa, Charmousis:2011bf}, or even beyond \cite{Appleby:2012rx, Gleyzes:2014dya,Kobayashi:2015gga,Ohashi:2015fma}. The Galileon/Horndeski type models generalized ordinary scalar field theories by including higher derivative terms and/or nonminimal couplings in specific forms such that the equation of motion remains of second order, i.e., there are  no redundant dynamical degree of freedom. This interesting feature has been applied to many contexts in cosmology where scalar fields play an important role, such as inflation \cite{Kobayashi:2010cm}, dark energy \cite{Deffayet:2010qz}, and  bouncing cosmology \cite{Qiu:2011cy}, etc.
 
In standard Galileon/Horndeski theories, the action contains only higher derivative terms or nonminimal coupling terms at the linear order. However, there are many well-motivated cosmological models whose action have non-linear forms. A prime example is the Dirac-Born-Infeld (DBI) action. The DBI action describes the dynamics of an extended object that breaks spontaneously the Poincare invariance of a higher dimensional theory \cite{Gliozzi:2011hj}. Indeed, DBI action arises in the context of D-branes in string theory  \cite{Polchinski:1998rq} (see e.g., \cite{Gerasimov:2000zp, Kutasov:2000qp,Kutasov:2000aq} for the derivation for some D-brane systems). When applied to inflation as in \cite{Silverstein:2003hf, Alishahiha:2004eh}, the inflaton field denotes the position of a D3-brane in a higher dimensional space. Aside from its fundamental origin, the square root feature of the DBI action brings several observational novelties. The most important of which is that the level of non-Gaussianities is enhanced by the inverse sound speed \cite{Chen:2006nt}. The observational signatures and the leading non-Gaussianities (bispectrum) of this generalized class of single field inflation models have been thoroughly examined in \cite{Chen:2006nt}, and extended to the inflationary trispectra in \cite{Huang:2006eha,Chen:2009bc}.
The strongest constraints on DBI inflation, especially on its sound speed, from the recent PLANCK data \cite{Ade:2015lrj}, come from the absence of such non-Gaussian features.
Such non-Gaussianity constraints also apply to other variants of DBI inflation, such as the IR 
model \cite{Chen:2004gc}, generalizations to other warped throats \cite{Kecskemeti:2006cg,Shiu:2006kj} and DBI inflation with multi-fields \cite{Huang:2007hh}.

As the DBI action is only an effective theory, it is subject to higher order corrections from the full theory of gravity. There have been many studies on the possible corrections to DBI actions in string theory, see e.g. \cite{Bachas:1999um,Wyllard:2000qe, Fotopoulos:2001pt, Wyllard:2001ye, Junghans:2014zla}. Moreover, the cases of putting gravitational terms into square-root has also been investigated for a long time in modified gravity theories, from Eddington in 1924 \cite{Eddington} to more recent literatures \cite{Deser:1998rj, Banados:2010ix}. Therefore in this paper, we discuss about another kind of correction, that is to have the kinetic term in the square-root of the action coupled to Einstein tensor. This coupling term can be found in Horndeski theory and shares some of its nice properties as mentioned above, but what is different is that now this term will appear as a non-linear term in the action, which may bring new interesting results. 
%For now we cannot tell how this term can be derived from fundamental theories because of the lack of the knowledge about them, however, assuming that this term does exist, phenomenologically it might have some interesting results, which we will explore in the following sections.

\section{Our new DBI model}
\subsection{background}
The action of our model is as follows:
\be\label{action}
S=\int d^{4}x\sqrt{-g}[\frac{R}{2\kappa^{2}}-\frac{1}{f(\phi)}(\sqrt{\cal D}-1)-V(\phi)]~,
\ee
where ${\cal D}\equiv1-2\alpha f(\phi)X+2\beta f(\phi)\tilde{X}$, and
\bea
\label{X}
X&\equiv&-\frac{1}{2}g^{\mu\nu}\nabla_{\mu}\phi\nabla_{\nu}\phi~,\nonumber\\
\label{tildeX}
\tilde{X}&\equiv&-\frac{1}{2M^2}G^{\mu\nu}\nabla_{\mu}\phi\nabla_{\nu}\phi~
\eea
where $M$ is some energy scale, therefore under FRW metric $g_{\mu\nu}=diag\{-1,a^2(t),a^2(t),a^2(t)\}$ where $a(t)$ is the scale factor of the universe, it becomes ${\cal D}=1-\alpha f(\phi)\dot\phi^2-3H^2\beta f(\phi)\dot\phi^2/M^2$. The form of $\tilde{X}$ has been proposed in \cite{Amendola:1993uh} and is widely studied in cosmology, see  \cite{Capozziello:1999xt, Capozziello:1999uwa, deRham:2011by, Banijamali:2012kq, Chen:2010ru, Feng:2013pba}. Since in the above action, the NKC term (namely $2\beta f(\phi)\tilde{X}$, or $\beta$-term in the following) which resides inside a square root is non-linear, the action cannot be transformed to that of the Galileon. More general actions with this feature have also been studied in \cite{Appleby:2012rx}, where the main focus is on the late-time acceleration. From our action eq.~(\ref{action}), one can straightforwardly obtain the equation of motion for $\phi$:
\bea\label{eomcov}
0&=&\frac{f_{\phi}}{f^{2}}(\sqrt{{\cal D}}-1)-\frac{f_{\phi}({\cal D}-1)}{2f^{2}\sqrt{\cal D}}-\frac{f_{\phi}}{2f^{2}{\cal D}^{3/2}}({\cal D}-1)^{2}~\nonumber\\
&&+\frac{f}{{\cal D}^{3/2}}(\alpha g^{\mu\nu}-\beta G^{\mu\nu})(\alpha\nabla_{\mu}X\nabla_{\nu}\phi-\beta\nabla_{\mu}\tilde{X}\nabla_{\nu}\phi)~\nonumber\\
&&+\frac{1}{\sqrt{\cal D}}(\alpha g^{\mu\nu}-\beta G^{\mu\nu})(\nabla_{\mu}\nabla_{\nu}\phi)-V_{\phi} 
\eea
and the Einstein Equation:
\bea\label{ee}
G_{\mu\nu}&=&\kappa^2T_{\mu\nu}~,\nonumber\\
T_{\mu\nu}&=&g_{\mu\nu}(-\frac{1}{f(\phi)}(\sqrt{\cal D}-1)-V(\phi))+\frac{\alpha\partial_{\mu}\phi\partial_{\nu}\phi}{\sqrt{\cal D}}~\nonumber\\
&&-2\frac{\beta R_{\rho\mu}\partial^{\rho}\phi\partial_{\nu}\phi}{\sqrt{\cal D}}+\frac{\beta\partial_{\mu}\phi\partial_{\nu}\phi}{2\sqrt{\cal D}}R-\frac{\beta X}{\sqrt{\cal D}}R_{\mu\nu}~\nonumber\\ 
&&+(\frac{\beta\partial_{\mu}\phi\partial^{\varepsilon}\phi}{\sqrt{\cal D}})_{;\nu;\varepsilon}+(\frac{\beta X}{\sqrt{\cal D}})_{;\nu;\mu}-\Box(\frac{\beta\partial_{\mu}\phi\partial_{\nu}\phi}{2\sqrt{\cal D}})~\nonumber\\
&&-2\Box(\frac{\beta X}{2\sqrt{\cal D}})g_{\mu\nu}-(\frac{\beta\partial^{\lambda}\phi\partial^{\varepsilon}\phi}{2\sqrt{\cal D}})_{;\varepsilon;\lambda}g_{\mu\nu}
\eea
Under FRW metric, Eqs. (\ref{eomcov}) becomes:
\bea\label{eom}
0&=&\frac{\frac{3\beta H^{2}}{M^2}+\alpha}{\mathcal{D}^{3/2}}(\ddot{\phi}+3H\mathcal{D}\dot{\phi})-\frac{f_{\phi}}{2f^{2}}(\frac{3\mathcal{D}-1}{\mathcal{D}^{3/2}}-2)~\nonumber\\
&&+\frac{3\beta}{M^2}H\dot{H}\frac{\mathcal{D}+1}{\sqrt{\mathcal{D}}}\dot{\phi}+V_{\phi}~,
\eea
and the energy density and pressure of the model can also be obtained by Eq. (\ref{ee}):
\bea
\label{rho}
\rho&=&\frac{(\sqrt{\cal D}-1)}{f(\phi)}+V(\phi)+\frac{\alpha\dot{\phi}^{2}}{\sqrt{\mathcal{D}}}+\frac{6\beta H^{2}\dot{\phi}^{2}}{M^2\sqrt{\mathcal{D}}}~,\\
\label{p}
p&=&-\frac{(\sqrt{\cal D}-1)}{f(\phi)}-V(\phi)-\frac{3\beta H^{2}\dot{\phi}^{2}}{M^2\sqrt{\mathcal{D}}}-(\frac{\beta H\dot{\phi}^{2}}{M^2\sqrt{\mathcal{D}}})^\cdot~,
\eea
where a `dot' denotes the time derivative with respect to time in Eq.s (\ref{eom}-\ref{p}). From the above one can see that in the limit of $\beta\rightarrow 0$ which implies the absence of the $\beta$-term, the action reduces to that of the usual DBI inflation. 
%However, with the $\beta$-term involved, things become different. One can already see from the equation (\ref{eom}) that the acceleration of $\phi$ now is determined not only by Hubble parameter $H$, but also by its derivative $\dot H$. Moreover, the pressure of the action also contains the second derivative of the field $\phi$. Actually, since $\rho$ and $p$ satisfy the friedmann equations:
%\be
%3M_p^2H^2=\rho~,~-2M_p^2\dot H=\rho+p~.
%\ee
%one could obtain the dynamical equation of $H$ as:
%\bea\label{friedmann2}
%&&(2-\frac{\beta\dot{\phi}^{2}}{M^2\sqrt{\mathcal{D}}}-\frac{3\beta^{2}H^{2}f\dot{\phi}^{4}}{M^4\mathcal{D}^{3/2}})\dot{H}\nonumber\\
%&=&-\frac{1}{\sqrt{\mathcal{D}}}(\frac{3\beta H^{2}}{M^2}+\alpha+\frac{f_{\phi}}{f}\frac{\beta(\mathcal{D}-1)}{2M^2\mathcal{D}}H\dot{\phi})\dot{\phi}^{2}\nonumber\\ &&+H\frac{\beta(\mathcal{D}+1)}{M^2\mathcal{D}^{3/2}}\dot{\phi}\ddot{\phi}~.
%\eea
%where $\ddot\phi$ get involved. This implies that the automatic system ${\phi,\dot\phi,H}$ now becomes a complicated system, where the dynamics of geometry and field interferes each other.
%In the limit of $\beta\rightarrow 0$, it is obvious to see that Eq.s (\ref{eom}) and (\ref{friedmann2}) reduce to those for the normal DBI inflation. 
For the non-slow-roll case ($\alpha f\dot\phi^2\lesssim 1$, or $\sqrt{\cal D}\gtrsim 0$, where $\sqrt{\cal D}$ is identified with the sound speed), large non-Gaussianity can be generated \cite{Chen:2006nt}, a feature disfavored by the latest Planck 2015 data \cite{Ade:2015lrj}. In the slow-roll limit ($\alpha f\dot\phi^2\ll 1$, $\sqrt{\cal D}\sim 1$), the action reduces to that of canonical scalar field inflation. 

However, when the $\beta$-term is important, there are interesting modifications, even in the slow-roll regime. One of which, as will be seen, is that the mass scale of the inflaton can be altered. We are interested in the case that the $\beta$-term dominated over the $\alpha$-term. In an FRW background we have $X\simeq \dot\phi^2/2$, $\tilde{X}\simeq-(H/M)^2\dot\phi^2/2$ respectively. The $\beta$ term dominates under the condition $H\gg M$. Taking the potential to be $V(\phi)=m^2\phi^2/2$ as an example where $m$ is the mass of the inflaton, the equation of motion (\ref{eom}) can be reduced as:
\be
\ddot\phi+(3-\epsilon-s)H\dot\phi-\left(\frac{mM}{H}\right)^2\phi=0~,
\ee
in the leading order. Here we have defined $\epsilon\equiv-\dot H/H^2$ and $s\equiv\dot{\sqrt{\mathcal{D}}}/(H\sqrt{\mathcal{D}})$. See also other slow-roll inflation models with NKC term in e.g. Ref. \cite{Capozziello:1999xt}. One can see that the effective mass of $\phi$ is $m_{eff}=mM/H$, i.e., it is suppressed by a factor $H/M$. This means the constraints on the mass of inflaton can be relaxed. Even a larger massive parameter $m$ can lead to the desired inflation scale $m_{eff}$.
%This property seems similar to that of canonical scalar field inflation with nominimal derivative coupling, however, as we will see later, they are still quite different in perturbation level, due to the nonlinearity of the action with respect to the Einstein curvature term.

\subsection{Perturbations} 
In addition to the inflationary background, it is also important to analyze its perturbations, for the second order perturbations seeds fluctuations in the Cosmic Microwave Background Radiation, as well as the galaxy distribution and gravitational waves. The perturbed line element can be written as follows:
\be\label{adm}
ds^{2}=-N^{2}dt^{2}+h_{ij}(dx^{i}+N^{i}dt)(dx^{j}+N^{j}dt)~,
\ee
where the lapse function $N$, the shift vector $N^i$ and the induced 3-metric $h_{ij}$ can be perturbed as:
\be\label{perturb}
N=1+A~,~N_i=\partial_i\psi~,~h_{ij}=a^2(t)e^{2\zeta+2\gamma_{ij}}~,
\ee
respectively. According to the ADM formalism, one can also decompose the elements in action (\ref{action}) as \cite{Wald:1984rg} (also see the Appendix in the first paper of \cite{Feng:2013pba}):
\bea\label{decompose}
R&=&^{(3)}R+K_{ij}K^{ij}-K^{2}~,~X=\dot\phi^2/2N~,\nonumber\\
\tilde{X}&=&-\frac{1}{4M^2}\frac{\dot{\phi}^{2}}{N^{2}}({}^{(3)}R-K{}_{ij}K^{ij}+K^{2})~,
\eea
here $^{(3)}R$ is the 3-curvature, and $K_{ij}$ is the extrinsic curvature, with $K\equiv Tr(K^i_j)$. Therefore, the second order action for tensor perturbation reads:
\be
S_2^T=\frac{1}{8\kappa^{2}}\int d^{4}xa^{3}[{\cal F}_T\dot{\gamma}_{ij}^2-{\cal G}_T\frac{(\nabla\gamma_{ij})^2}{a^{2}}]~.
\ee
from the action one can get the sound speed squared for the tensor perturbations:
\be\label{cT2}
{\cal F}_T=1-\frac{\kappa^{2}\beta\dot{\phi}^{2}}{2M^2\sqrt{\mathcal{D}}}~,~{\cal G}_T=1+\frac{\kappa^{2}\beta\dot{\phi}^{2}}{2M^2\sqrt{\mathcal{D}}}~,~c_{T}^{2}\equiv\frac{{\cal G}_T}{{\cal F}_T}~.
\ee

In order to maintain stability under tensor perturbation, the sound speed squared must be positive, i.e., $c_T^2>0$. From (\ref{cT2}) this implies $2M^2\sqrt{\mathcal{D}}>\kappa^{2}|\beta|\dot{\phi}^{2}$, which will give us a constraint in the following analysis. Note that in the ``slow-roll" limit where $\kappa^{2}|\beta|\dot{\phi}^{2}\ll2M^2\sqrt{\mathcal{D}}$, $c_T^2$ is approaching  unity, recovering the standard case. From the action, one gets the equation of motion for $\gamma$:
\be
\gamma_{ij}^{\prime\prime}-c_T^2\nabla^2\gamma_{ij}+\frac{(a^2{\cal F}_T)^\prime}{a^2{\cal F}_T}\gamma_{ij}^\prime=0~,
\ee
which has the following solutions:
\be
\gamma_{ij}=\text{constant.},~~~~\int\frac{dt}{a^3(t){\cal F}_T}~,
\ee
The power spectrum for tensor perturbations is therefore:
\be\label{tensorspectrum}
P_{T}\equiv\frac{k^3}{2\pi^2}|\gamma_{ij}|^2=\frac{2H^{2}}{{\cal G}_T c_{T}\pi^{2}}~,
\ee
and the spectral index
\be
n_{T}\equiv\frac{d\ln P_T}{d\ln k}=\frac{\kappa^{2}\beta\dot{\phi}^{2}}{\kappa^{2}\beta\dot{\phi}^{2}-2M^2\sqrt{\mathcal{D}}}(2\iota-s)-2\epsilon-s_{T}
\ee
where $s_{T}\equiv\dot{c}_{T}/(Hc_{T})$, $\iota\equiv\ddot\phi/(H\dot\phi)$. From those results we find that the power spectrum for gravitational waves deviates from scale invariance only up to slow-roll corrections. However, its difference from the standard minimal coupling case is also of the order of the slow-roll parameters, so if the sensitivity of the future observations is smaller than the slow roll parameters, we can distinguish our model from inflation models with minimal coupling.

What is more promising observationally is the feature in scalar perturbations. Substituting Eqs.(\ref{perturb}) and (\ref{decompose}) into action (\ref{action}) and taking the scalar parts, one finds the scalar perturbation action. First of all, considering the Hamilton and momentum constraint equations: (1) $\delta S/\delta N=0$; (2) $\delta S/\delta N_i=0$, we obtained the following two equations:
\bea\label{constraint1}
0&=&-[12H^{2}+\frac{2\kappa^{2}}{f(\phi)\sqrt{\mathcal{D}}}(1-\frac{9\beta H^{2}}{M^2}f(\phi)\dot{\phi}^{2})\nonumber\\
&&-\frac{2\kappa^{2}}{f(\phi)\mathcal{D}^{3/2}}(1+\frac{3\beta H^{2}}{M^2}f(\phi)\dot{\phi}^{2})^{2}]A+12H[1\nonumber\\
&&-\frac{\kappa^{2}\beta\dot{\phi}^{2}}{M^2\sqrt{\mathcal{D}}}-\frac{\kappa^{2}\beta\dot{\phi}^{2}}{2M^2\mathcal{D}^{3/2}}(1+\frac{3\beta H^{2}}{M^2} f(\phi)\dot{\phi}^{2})]\dot{\zeta}-4a^{-2}[1\nonumber\\
&&-\frac{\kappa^{2}\beta\dot{\phi}^{2}}{2M^2\mathcal{D}^{3/2}}(1+\frac{3\beta H^{2}}{M^2} f(\phi)\dot{\phi}^{2})]\partial^{2}\zeta-4a^{-2}H[1\nonumber\\
&&-\frac{\kappa^{2}\beta\dot{\phi}^{2}}{M^2\sqrt{\mathcal{D}}}-\frac{\kappa^{2}\beta\dot{\phi}^{2}}{2M^2\mathcal{D}^{3/2}}(1+\frac{3\beta H^{2}}{M^2} f(\phi)\dot{\phi}^{2})]\partial^{2}\psi~,
\eea
\bea
\label{constraint2}
0&=&[2(1-\frac{\kappa^{2}\beta\dot{\phi}^{2}}{M^2\sqrt{\mathcal{D}}})H-\frac{\kappa^{2}\beta H\dot{\phi}^{2}}{M^2\mathcal{D}^{3/2}}(1+\frac{3\beta H^{2}}{M^2} f(\phi)\dot{\phi}^{2})]A\nonumber\\
&&-(2-\frac{\kappa^{2}\beta\dot{\phi}^{2}}{M^2\sqrt{\mathcal{D}}})\dot{\zeta}+\frac{\kappa^{2}\beta^{2}Hf(\phi)\dot{\phi}^{4}}{M^4\mathcal{D}^{3/2}}(3H\dot{\zeta}-a^{-2}\partial^{2}\zeta\nonumber\\
&&-a^{-2}H\partial^{2}\psi)~.
\eea

Interestingly, in the second (momentum) constraint equation, there are not only terms involving $A$ and $\dot\zeta$, but also nontrivial terms in $\partial^2\zeta$ and $\partial^2\psi$, which originate from the $\beta$-term. In the usual minimal coupling case where these terms are absent, one can easily find the simple relations $A\sim\dot\zeta$, $\partial^2(\zeta+H\psi)\sim\dot\zeta$ which, upon substituting into the action (\ref{action}) result in a second order perturbation action that is homogeneous in the order of derivatives \cite{Chen:2006nt}. However,  it is not the case here. It is impossible to cancel those extra terms, and combining (\ref{constraint1}) and (\ref{constraint2}) will give solutions to $A$ and $\partial^2\psi$ which both inevitably contain $\partial^2\zeta$ and $\dot\zeta$, making the perturbed action non-quadratic in the spatial derivatives. This is due to the fact that our action is no longer linear in $\tilde{X}$ (given in eq.~(\ref{tildeX})), or more essentially, the extrinsic curvature $K_{ij}K^{ij}-K^{2}$ which is related to $\tilde{X}$ through eq.~(\ref{decompose}). In the standard case where the action is linear in $K_{ij}K^{ij}-K^{2}$, such as $S\sim\Upsilon(K_{ij}K^{ij}-K^{2})$ where $\Upsilon$ is a space-independent background parameter, variation of the action w.r.t. $\delta N_i$ gives rise to $\Upsilon\nabla_i(K_{ij}K^{ij}-K^{2})=\Upsilon(HA-\dot\zeta)=0$, leading to $A=\dot\zeta/H$. However, for a nonlinear action, $\Upsilon$ also contains space-dependent terms. Therefore the second constraint equation becomes $\nabla_i[\Upsilon(K_{ij}K^{ij}-K^{2})]=\Upsilon\nabla_i(K_{ij}K^{ij}-K^{2})+(K_{ij}K^{ij}-K^{2})\nabla_i\Upsilon=0$, and nontrivial terms from $\nabla_i\Upsilon$ appear. That is why we get correction terms such as $\partial^2\zeta$ and $\partial^2\psi$ which, as will be shown below, will modify the dispersion relations of the perturbation.

In principle, following the way above, one can construct  even more variants of the DBI action such that the constraint equations get modified, by adding nonlinear gravitational terms such as $R^n$, $R_{\mu\nu}G^{\mu\nu}$, or $R^{...}_{...}R^{...}_{...}$ etc. However, as is well-known, most of these terms will introduce higher derivatives in the background equations of motion (\ref{eom}). As a result, new dynamical degrees of freedom need to be introduced. In many cases, these higher derivative terms will lead to ghost instability, however, in our model the background equations of motion remain second order as in the Galileon theories, so there would be no ghost, which will be shown below (the same argument has also been presented in \cite{Appleby:2012rx}). 

%\footnote{In some modified gravity theories like $f(R)$ or $f(\cal G)$, where $\cal G$ is the Gauss-Bonnet term, there is no scalar field, the uniform gauge becomes $\delta R=0$ or $\delta{\cal G}=0$, or equivalently $\delta f_{,R}(f_{,{\cal G}})=0$. This is because we can only construct gauge invariant perturbation variable with $\delta f_{,R}$ or $\delta f_{,{\cal G}}$ \cite{deFeliceTsujikawa}. However, when such condition is imposed, the correction term will disappear and dispersion relation will recover the quadratic one. This implies that these models are equivalent to the single scalar model. However, in our case we also have the scalar d.o.f. $\phi$, so our model is actually a nonminimal coupling model with nonlinear coupling term. So it is different.}.

The constraint perturbation variables $A$ and $\partial^2\psi$ can in principle be solved, however the expressions are complicated, and thus not useful for our analysis. For illustrative purposes, we will be content with some limits and  consider only the leading order terms. First of all, it is quite useful to define several dimensionless variables, which will bring convenience to our calculations:
\be
x_\beta\equiv\frac{\kappa^{2}\beta\dot{\phi}^{2}}{2M^2\sqrt{\mathcal{D}}}~,~x_\alpha\equiv\frac{\alpha\dot{\phi}^{2}}{2\sqrt{\mathcal{D}}}~,~y\equiv \frac{f(\phi)M_{p}^2H^{2}}{\sqrt{\mathcal{D}}}~.
\ee
The positivity of the sound speed for the tensor perturbation  requires $2M^2\sqrt{\mathcal{D}}>\kappa^{2}|\beta|\dot{\phi}^{2}$ which leads to the condition $|x_\beta|<1$. Therefore we will consider the limit of $|x_\beta|\ll 1$ for illustrative purposes. Moreover, $|x_\alpha|\ll 1$ is also required by the slow-roll condition. With the help of these variables, Eqs.~(\ref{constraint1}) and (\ref{constraint2}) can be approximated as:
\bea
0&\simeq&[6H^{2}-\frac{\kappa^{2}(1-\mathcal{D})}{f(\phi)\mathcal{D}^{3/2}}]A-6H\dot{\zeta}+\frac{2}{a^{2}}\partial^{2}\zeta\nonumber\\
&&+\frac{2H}{a^{2}}\partial^{2}\psi~,\\
0&\simeq&HA-\dot{\zeta}-\frac{2x_\beta^{2}y}{a^{2}H}\partial^{2}\zeta-\frac{2x_\beta^{2}y}{a^{2}}\partial^{2}\psi~,
\eea
which have the solution:
\be
A\approx\frac{\dot{\zeta}}{H}-\frac{4x_\beta^{3}y}{a^{2}H^{2}}\partial^{2}\zeta~,
~\partial^2\psi\approx\frac{a^{2}(1-\mathcal{D})}{2\mathcal{D}^{2}y}\dot{\zeta}-\frac{1}{H}\partial^{2}\zeta~.
\ee

Substituting into the second order perturbed action, one finds
%\bea\label{perturbactionscalar}
%S_2^c&\approx&\frac{1}{2\kappa^{2}}\int d^{4}xa^{3}\Big[\frac{(1-\mathcal{D})}{\mathcal{D}^{2}y}\dot{\zeta}^{2}-\frac{2\epsilon}{a^{2}}(\partial\zeta)^{2}\nonumber\\ &&+\frac{16x_\beta^{4}y}{a^{4}H^{2}}(\partial^{2}\zeta)^{2}\Big]~.
%\eea
%Note that in the absence of the $\beta$ term, the factor in front of $\dot{\zeta}^{2}$ reduces to $(1-\mathcal{D})/\mathcal{D}^{2}y=\kappa^2\alpha\dot\phi^2/(H^2{\cal D}^{3/2})=\epsilon/c_s^2$, which is consistent with \cite{Chen:2006nt}. However, in the presence of the $\beta$-term, $\cal D$ can be rewritten as ${\cal D}=1-2\mathcal{D}(x_\alpha y+3x_\beta y)$, and $(1-\mathcal{D})/\mathcal{D}^{2}y=2(x_\alpha+3x_\beta)/{\cal D}$. Since we are here interested in the $H\gg M$ case where the $\beta$-term dominates over the $\alpha$-term, the perturbed action (\ref{perturbactionscalar}) then becomes:
\bea\label{perturbactionscalar2}
S_2^c&\approx&\frac{1}{2\kappa^{2}}\int d^{4}xa^{3}\Big[6\frac{x_\beta}{\mathcal{D}}\dot{\zeta}^{2}-\frac{2\epsilon}{a^{2}}(\partial\zeta)^{2}\nonumber\\ &&+\frac{16x_\beta^{4}y}{a^{4}H^{2}}(\partial^{2}\zeta)^{2}\Big]~,
\eea
where we've made use of the assumption of $H\gg M$ and that the $\beta$-term dominates over the $\alpha$-term. Note that in order to get rid of the ghost problem, only $x_\beta>0$ is allowed. This means that we need a positive coupling constant $\beta$ in this new DBI inflation model. Moreover, the action (\ref{perturbactionscalar2}) makes clear that in addition to the normal quadratic terms, there is also a quartic term $(\partial^2\zeta)^2$. This means that in the large $k$ limit, the dispersion relation will get modified. We will further discuss this novelty below.

From the action in eq.~(\ref{perturbactionscalar2}), one gets the equation of motion for the curvature perturbation $\zeta$ in a canonical form:
\be\label{perturbeomscalar2}
u^{\prime\prime}+\omega^2u-\frac{z^{\prime\prime}}{z}u=0~,
\ee
where $u\equiv z\zeta$, $z\equiv a\sqrt{3x_\beta/{\cal D}}$, prime denotes derivative with respect to conformal time: $\eta\equiv\int a^{-1}(t) dt$, and 
\be\label{omega2}
\omega^2=\frac{\epsilon{\cal D}}{3x_\beta}k^2\left[1+24\frac{x_\beta^5|y|}{\epsilon^2{\cal D}^2}\left(\frac{c_sk}{aH}\right)^2\right]~.
\ee
with $c_s^2=\epsilon{\cal D}/3x_\beta$. We also made use of $\partial\rightarrow ik$ to transfer this equation into the Fourier space. 

From Eq. (\ref{omega2}) one can see that, the fluctuation modes can be divided into large-scale ones $(k<k_c)$ and small-scale ones $(k>k_c)$ by a critical $k$-value: $k_c\equiv aH\sqrt{\epsilon{\cal D}/(8x_\beta^4y)}$, where the approximate dispersion relation approaches $\omega\sim k^2\eta$ and $\omega\sim k$ respectively. Here we further made use of the relation $aH\sim-1/\eta$.
%\be\label{perturbeomscalar}
%\zeta^{\prime\prime}+c_s^2k^2\zeta+\frac{(a^2x_\beta/{\cal D})^\prime}{(a^2x_\beta/{\cal D})}\zeta^\prime-\frac{8|x_\beta|^3y}{3a^2H^2}k^4\zeta=0~,
%\ee
Since $a$ is growing with time while the other parameters are nearly constant, $k_c$ is also time-growing. Therefore, the wavenumber $k$ of those large-scale modes which is less than $k_c$ at the initial time of inflation will never exceed $k_c$, and the $k^4$-term will never enter into the solution. Conversely, for the small-scale modes with $k$ initially larger than $k_c$, $k^4$-term can have significant effects at the beginning of inflation.

Solving Eq. (\ref{perturbeomscalar2}) we  get:
\be
u=\frac{\sqrt{\pi|\eta|}}{2}[H^{(1)}_{\nu}(\omega\eta)+H^{(1)}_{-\nu}(\omega\eta)]~,
\ee
where $H^{(1)}$ is the type I Hankel function, and we approximately have $\nu\simeq 3\int\omega d\eta/(2\omega\eta)$. For large-scale modes we have $\nu\simeq 3/2$, while for small-scale modes one has $\nu\simeq 3/4$. Therefore for the large-scale modes where the $k^4$-term can be neglected, making use of $u=z\zeta$, one gets the standard solution:
\bea
\zeta&\simeq&\frac{1}{\sqrt{2k}}e^{ik\eta}~,~~~~~~~~~~~~~~~~~~~~~~~(subhorizon)~\nonumber\\
&\sim&\text{constant.},~~\frac{1}{3}\int\frac{{\cal D}dt}{a^3(t)x_\beta}~,~(superhorizon)~
\eea
where the subhorizon solution is the standard Bunch-Davies vacuum solution, while in the superhorizon solution the first branch is the dominant one. The power spectrum and spectral index are then:
\bea
P^{(l)}_S&\equiv&\frac{k^3}{2\pi^2}|\zeta|^2\simeq\frac{H^2}{8\pi^2}\sqrt{\frac{3x_\beta}{\epsilon^3{\cal D}}}~,\\
n^{(l)}_S&\equiv&1+\frac{d\ln P^{(l)}_S}{d\ln k}\simeq1+2\epsilon-\frac{3}{2}\eta_e+\iota-\frac{3}{4}s~.
\eea
where $\eta_e\equiv\dot\epsilon/H\epsilon$. The index ``$(l)$" denotes ``large scales". Together with the tensor spectrum (\ref{tensorspectrum}), one can also get the tensor-to-scalar ratio:
\be
r^{(l)}\equiv\frac{P_T}{P^{(l)}_S}\simeq 16\epsilon\sqrt{\frac{\epsilon{\cal D}}{3x_\beta}}~
\ee
in its leading order for large-scale modes. 
   
For the small-scale modes where the $k^4$-term has to be taken into account, the approximate solution of $\zeta$ in the outside-horizon region is
\bea
\zeta&\simeq&\frac{1}{\sqrt{2\omega(k,\eta)}}\exp\left(i\int^\eta \omega(k,\eta^\prime)d\eta^\prime\right)~,~(subhorizon)~\nonumber\\
&\sim&H\sqrt{\frac{{\cal D}}{6x_\beta\omega^3}}~,~~H\sqrt{\frac{\omega^3{\cal D}}{6x_\beta}}|\eta|^3~,~(superhorizon)~
\eea
where the subhorizon solution reduces to WKB solutions, which is consistent with Ref. \cite{Cai:2009hc, Lu:2009he} (see also \cite{DeFelice:2014bma}), while in the superhorizon solution the first branch is the dominant one. 

The power spectrum is:
\be
P^{(s)}_S\simeq\frac{H^2}{8\pi^2}\sqrt{\frac{3x_\beta}{\epsilon^3{\cal D}}}\left[1-36\frac{x_\beta^5|y|}{\epsilon^2{\cal D}^2}\left(\frac{c_sk}{aH}\right)^2\right]~,
\ee
and the index ``$(s)$" means ``small scales". At the crossing time when $c_sk\simeq aH$, the term inside the round brackets reduces to 1. This shows that due to the $k^4$-term, there is a deficit at small scales in the power spectrum, which can cause an additional red-tilt of the spectrum, aside from the one due to the running of the parameters. This is a new effect of our model.

The spectral index is:
\bea
n^{(s)}_S&\simeq&1+\frac{d\ln P^{(s)}_S}{d\ln k}=1+2\epsilon-\frac{3}{2}\eta_e+\iota-\frac{3}{4}s+\Delta n~,\\
\Delta n&=&-\frac{36x_\beta^5|y|/\epsilon^2{\cal D}^2}{1-36x_\beta^5|y|/\epsilon^2{\cal D}^2}(5\epsilon_x+\epsilon_y-2\eta-2s)~,
\eea
where $\epsilon_x\equiv\dot{x}_\beta/Hx_\beta$, $\epsilon_y\equiv\dot y/Hy$. One can see that, the $k^4$-term only gives rise to a correction of the order of the slow-roll parameters (or even smaller) to the spectral index, so nearly scale invariance will not be spoiled. The reason for this is that the $k^4$-term also depends on $a$ and $H$, and furthermore take the form of  $k^2(k/aH)^n$ where $n$ is an arbitrary integer. Corrections of this form will not alter the scale-invariance of the power-spectrum, as has been proved in the general form in Ref. \cite{Cai:2009hc, Lu:2009he}.

Finally, the tensor-to-scalar ratio reads
\be
r^{(s)}\equiv\frac{P_T}{P^{(s)}_S}\simeq 16\epsilon\sqrt{\frac{\epsilon{\cal D}}{3x_\beta}}\left[1+36\frac{x_\beta^5|y|}{\epsilon^2{\cal D}^2}\left(\frac{c_sk}{aH}\right)^2\right]~
\ee
for small-scale modes.

\subsection{Constraints on parameters}
From the results above, it is possible to constrain the parameters of our model with observational data. In our model, there are two critical parameters, $x_\beta$ and $y$. The first one corresponds to the strength of the NKC, while both of them determines the largeness of $k^4$ correction. 

From the result for the tensor-to-scalar ratio, it is straightforward to express $x_\beta$ as:
\be
x_\beta\simeq8.53\times10^{-3}\times\left(\frac{\epsilon}{0.01}\right)^3\left(\frac{0.1}{r}\right)^2~,
\ee
which means, if future observations confirm that $r$ is of ${\cal O}(0.1)$ and the slow-roll parameter $\epsilon\simeq 0.01$, then $x_\beta$ will be of order $10^{-2}$ to $10^{-3}$, namely the same order with $\epsilon$, which is consistent with our analysis above. Furthermore, from the small-scale power spectrum one finds:
\be
y\simeq\frac{\Delta P_S}{P_S}\left(\frac{\epsilon}{x_\beta}\right)^5\left(\frac{0.01}{\epsilon}\right)^3\times10^6~.
\ee
where $\Delta P_S/P_S$ is the relative difference of the power spectra due to the corrections. If the future observations can distinguish different sources for this difference, our model can be tested. An observation of step-like variation of the spectra tends to support our model, where the parameter $y$ (which is related to the function $f(\phi)$ in action (\ref{action})) can be determined. 

\section{Conclusion} 
In this paper, we studied a new DBI inflation model, with a NKC term under the square-root. Being nonlinear in the action, this term may give us interesting phenomena. On the background level, this term will contribute to the effective mass of the inflaton, so that the constraint on the inflaton mass can be relaxed.

On the perturbation level, our results are two-folded. On the one hand, even though the higher-order correction is presented in the perturbative equation (which is due to the non-linearity of the NKC term), the result is not altered significantly that a nearly scale-invariant power spectrum could still be obtained. This is because the correction term is in a special form, in which the dependences on $k$ can be self-canceled. On the other hand, the spectrum shows a deficit in the spectrum at small scales and gives rise to a red tilt, as a result of  the corrections in the equation of motion. 
%does change a little, in the amplitude. The observational data favors a red-tilted spectrum, which means a deficit in large scales. In our model, other than the usual smooth $k$-dependence which is presented by the deviation of $n_S$ from unity, there are also a deficit in the amplitude of the spectrum due to the higher order corrections. 
If future observations can distinguish this effect from the one caused by running of the slow-roll parameters, it may provide a test of our model. It would also be interesting to see if our action (\ref{action}) can be derived from breaking spacetime symmetries of some higher dimensional theory \cite{Hidaka:2014fra}, in a similar spirit as how the DBI action arises. 

%the quartic term will take action only for those $k$ modes with Note that in slow-roll limit we have $3|\beta f(\phi)|(H/M)^{2}\dot{\phi}^{2}\approx 3|\beta f(\phi)|(H/M)^{2}\dot{\phi}^{2}/\mathcal{D}=|x|[6|f(\phi)|(M_{p}H)^{2}/\mathcal{D}]\ll1$, so $6|x_\beta y|\ll1$, therefore corrections will only affect very large $k$-modes. We assume that those $k<k_c$ modes, where quartic term can be neglected, are responsible for the primordial power spectrum which can be observed in CMB. For those modes, one find the solution:
\begin{acknowledgments}
T.Q. thanks Yun-Song Piao and Gary Shiu for useful discussions. The work of T.Q. is supported  in part by NSFC under Grant No: 11405069, in part by the Open Project Program of State Key Laboratory of Theoretical Physics, Institute of Theoretical Physics, Chinese Academy of Sciences, China (No.Y4KF131CJ1), and in part by by the Open Innovation Fund of Key Laboratory of Quark and Lepton Physics (MOE), Central China Normal University (No.:QLPL2014P01).
\end{acknowledgments}

\end{document}